\documentclass[twocolumn]{aastex701}
\usepackage{amsmath}
\usepackage{breqn}
\usepackage{hyperref}
\usepackage{url}

\font\manual=manfnt at 7pt \def\dbend{\hbox{\raise0.9ex\hbox{\manual\char127\hspace{0.6em}}}}

\providecommand{\e}[1]{\ensuremath{\times 10^{#1}}}

\newcounter{INTERNALionstage}

\def\gtsim{\mathrel{\hbox{\rlap{\hbox{\lower4pt\hbox{$\sim$}}}\hbox{$>$}}}}
\def\lesssim{\mathrel{\hbox{\rlap{\hbox{\lower4pt\hbox{$\sim$}}}\hbox{$<$}}}}

\def\A{{\rm\thinspace \AA}}

\def\g{{\rm\thinspace g}}

\def\K{{\rm\thinspace K}}

\def\m{{\rm\thinspace m}}

\def\s{{\rm\thinspace s}}

\def\W{{\rm\thinspace W}}

\def\fexxv{\mbox{{\rm Fe~{\sc xxv}}}}
\def\fexxvi{\mbox{{\rm Fe~{\sc xxvi}}}}

\def\h0{\mbox{{\rm H}$^0$}}
\DeclareMathAlphabet{\vib}{OML}{cmm}{m}{it}

\renewcommand{\thefootnote}{\fnsymbol{footnote}}

\shorttitle{XRISM observations of A1795}
\graphicspath{{./}{figures/}}

\begin{document}

\title{
XRISM Observations of Abell 1795: Evidence for Low Turbulence and Resonant Scattering}

\author[0000-0002-5222-1337]{Arnab Sarkar}
\affiliation{Department of Physics, University of Arkansas, 825 W Dickson st.,
Fayetteville, AR 72701, USA}
\affiliation{Kavli Institute for Astrophysics and Space Research,
Massachusetts Institute of Technology, 70 Vassar St, Cambridge, MA 02139}
\email{arnabs@uark.edu}

\author[0000-0002-3031-2326]{Eric D.\ Miller}
\affiliation{Kavli Institute for Astrophysics and Space Research,
Massachusetts Institute of Technology, 70 Vassar St, Cambridge, MA 02139}
\email{}

\author{Brian McNamara}
\affiliation{Department of Physics \& Astronomy, Waterloo Centre for Astrophysics, University of Waterloo, Ontario N2L 3G1, Canada}
\email{}

\author{Helen Russell}
\affiliation{School of Physics \& Astronomy, University of Nottingham, University Park, Nottingham NG7 2RD, UK}
\email{}

\author[0000-0001-8055-7113]{Kotaro Fukushima}
\affiliation{Institute of Space and Astronautical Science (ISAS), Japan Aerospace Exploration Agency (JAXA), Kanagawa 252-5210, Japan}
\email{}

\author{Mark Bautz}
\affiliation{Kavli Institute for Astrophysics and Space Research,
Massachusetts Institute of Technology, 70 Vassar St, Cambridge, MA 02139}
\email{}

\author[0000-0003-0058-9719]{Yutaka Fujita}
\affiliation{Department of Physics, Tokyo Metropolitan University, Tokyo 192-0397, Japan}
\email{}

\author[0000-0002-4737-1373]{Catherine E. Grant}
\affiliation{Kavli Institute for Astrophysics and Space Research,
Massachusetts Institute of Technology, 70 Vassar St, Cambridge, MA 02139}
\email{}

\author[0000-0002-7031-4772]{Fran\c{c}ois Mernier}
\affiliation{Department of Astronomy, University of Maryland, College Park, MD 20742, USA} 
\affiliation{NASA / Goddard Space Flight Center, Greenbelt, MD 20771, USA}
\affiliation{Center for Research and Exploration in Space Science and Technology, NASA / GSFC (CRESST II), Greenbelt, MD 20771, USA}
\affiliation{Univ Toulouse, CNES, CNRS, IRAP, Toulouse, France}
\email{}

\author{Michael A. McDonald}
\affiliation{Kavli Institute for Astrophysics and Space Research,
Massachusetts Institute of Technology, 70 Vassar St, Cambridge, MA 02139}
\email{}

\author[0000-0002-2784-3652]{Naomi Ota}
\affiliation{Department of Physics, Nara Women's University, Nara 630-8506, Japan}
\email{}

\author{Ay\c{s}eg\"{u}l T\"{u}mer}
\affiliation{Center for Space Sciences and Technology, University of Maryland, Baltimore County (UMBC), Baltimore, MD, 21250 USA}
\affiliation{NASA / Goddard Space Flight Center, Greenbelt, MD 20771, USA}
\affiliation{Center for Research and Exploration in Space Science and Technology, NASA / GSFC (CRESST II), Greenbelt, MD 20771, USA}
\email{}

\author[0000-0001-9110-2245]{Daniel Wik}
\affiliation{Department of Physics \& Astronomy, University of Utah, 270 South 1400 East, Salt Lake City, UT, 84112, USA}
\email{}



\begin{abstract}
We present high-resolution X-ray spectroscopic observations of the cool-core galaxy cluster Abell~1795 obtained with XRISM/Resolve.
The cluster was observed with two deep pointings: a 225 ks central exposure and a 113 ks northern exposure, extending to a projected radius of 320 kpc from the cluster center. 
Single-temperature fits reveal a clear
radial gradient in the line-of-sight 
velocity dispersion, decreasing from
{ 114 $\pm$ 11 km/s in the core to
68 $\pm$ 39 km/s at 320 kpc. 
The bulk velocities in the central regions
are very low (22 $\pm$ 12 and 7 $\pm$ 21 km/s)},
indicating no significant relative motion
between the brightest cluster galaxy (BCG)
and the intracluster medium (ICM).
Given that the central region includes the 
southward-extending cool gas tail, this 
result disfavors the ``cooling-wake'' scenario
and instead supports an AGN-uplift origin.
We find that the nonthermal pressure fraction
decreases with radius, from 
$P_{\rm NT}/P_{\rm T}\approx2\%$ in the
core to $\sim0.6\%$ at 330 kpc, suggesting 
that the northern ICM of A1795 is largely
quiescent. 
Two-temperature and split energy-band
(2--4 keV and 6--7 keV) fits identify
two gas phases within the central $<1.5'$
region, providing strong evidence for
multiphase gas in the cluster core.
We detect a $\sim14\%$ resonant suppression
of the optically thick $\fexxv$ $w$ line in
the center. Additionally, we observe
a significant excess in the $\fexxv$ $y$ 
line-flux relative to models. Accounting for
uncertainties in the atomic data reduces this
discrepancy, suggesting that atomic
data uncertainties may contribute to the
observed residual flux.
\end{abstract}


\section{Introduction}
Galaxy clusters assemble 
hierarchically through mergers 
and ongoing accretion from the cosmic web.
The resulting gas dynamics play a key role
in the cluster energy budget, as 
a substantial fraction of the injected kinetic energy is transferred to the
intra-cluster medium (ICM), 
driving bulk motions 
and turbulence 
\citep[e.g.,][]{Vazza2009,Vazza2018,Simionescu2019,2022ApJ...935L..23S,2023ApJ...944..132S,2025ApJ...984L..63S}.
The first direct measurement of line-of-sight ICM velocities in a relaxed cluster core provided by Hitomi observations of Perseus indicates that, despite the rich spatial structure revealed by Chandra 
\citep{2007MNRAS.381.1381S,2016MNRAS.460.1898S}, the ICM is quiescent,
with velocity dispersion less than 
200 km/s within the central
100 kpc and non-thermal pressure support
of only $\sim$ 4\%. 
\citep[e.g.,][]{Hitomi2016,Hitomi2_2018}. 
These results suggest that the velocity broadening is driven by small-scale motions,
with a turbulent driving scale
less than 100 kpc, consistent with the 
size of Active Galactic Nuclei (AGN)-driven bubbles; that non-thermal pressure accounts for a small 
fraction of the total pressure in the core; 
and that sloshing can explain a modest bulk velocity gradient across the cluster core.

The launch of the XRISM satellite in 2023 has opened 
up a new opportunity to probe kinematics
in other clusters using an instrument with similar spectral 
capabilities to the Hitomi SXS \citep{2025PASJ...77S...1T,Kelley2025_ResolveJATIS}.
In the past year, the Resolve micro-calorimeter
onboard XRISM has revealed line-of-sight (LOS) velocity dispersion
measurements in several cool-core clusters such
as Abell~2029 
\citep{A2029_eric,A2029_naomi,A2029_Sarkar},
Centaurus \citep{2025Natur.638..365X}, 
Hydra A \citep{2025ApJ...990...42R},
and the Ophiuchus cluster \citep{2025PASJ...77S.270F}.
Their velocity dispersion estimates lie between
120--190 km/s. 
Even for merging galaxy clusters like Coma
\citep{2025ApJ...985L..20X}
and Abell~3667 
\citep{2026ApJ...996L..15O},
the velocity dispersion is
$\sim$200 km/s.
These findings, while groundbreaking, highlight our lack of understanding about how the effects of AGN feedback, 
merger-induced sloshing, and other dynamical activity conspire to shape the spatial 
and kinematic structure of clusters on all scales.
We can address this by dramatically expanding 
the sample of clusters with accurate direct velocity measurements.

\begin{figure*}
 \centering
\includegraphics[width=1\textwidth]{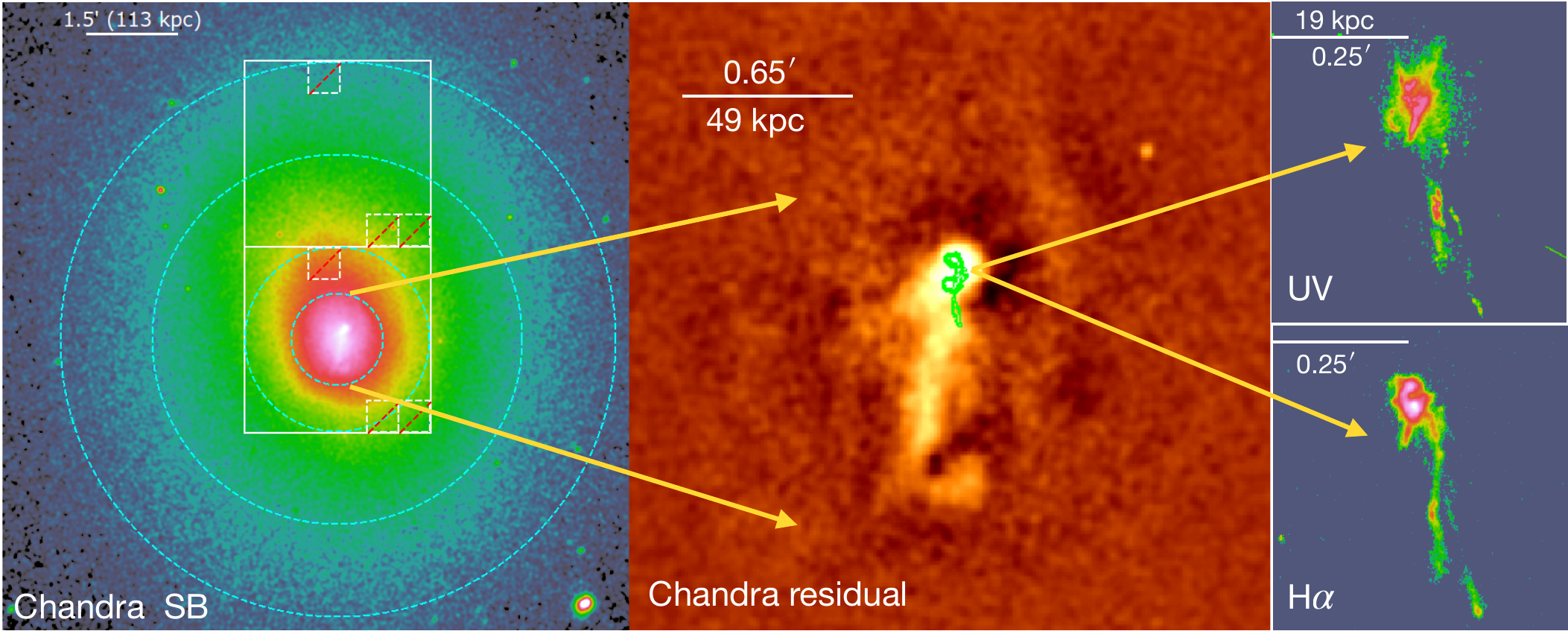}
    \caption{Left: Exposure corrected and background subtracted Chandra image of
    A1795 in the 0.5-10 keV energy band.
    Resolve FoVs are marked with white boxes. Dashed small white boxes show
    the excluded pixels 11, 12 (calibration pixel), and 27.
    Cyan dashed annuli shows the regions
    used for ARFs generation.
    Right: Zoomed in residual image of A1795 
    after
    subtracting a beta model image.
    Several substructures including the `hook' are clearly seen. Green contours represent
    the ALMA CO(2--1)
    contours from \citet{2017MNRAS.472.4024R}.
    }
    \label{fig:chandra_image}
\end{figure*}

Abell~1795 (A1795 hereafter) is a
nearby ($z=0.062$),
relatively hot 
(spatially-averaged $kT\approx5.3$ keV;
\citealt{Vikhlinin_2006,2008A&A...478..615S,2009PASJ...61.1117B}) cool-core
cluster. 
It is one of a handful of clusters
with very deep Chandra observations, 
exceeding 3.4 Msec, due to its use as a regular calibration source. 
Several studies have utilized this dataset and extensive multiwavelength coverage to paint a picture of a complex core of activity
within the inner 100 kpc (1.4$\arcmin$) as illustrated in Figure \ref{fig:chandra_image}
\citep[see also:][]{Walker2014,Ehlertetal2015,Kokotanekov2018_A1795}.
Multiple AGN-blown cavities are seen 
along with radio emission whose spectral signatures suggest several 
episodes of AGN activity. 
A 50-kpc-long tail of cool X-ray-emitting
plasma extends south of the Brightest Cluster Galaxy (BCG),
spatially coincident with striking H$\alpha$ filaments and ending in a hook-shaped 
feature 
\citep{McDonaldVeilleux2009,Ehlertetal2015}. 
This tail is thought to arise from a cooling 
wake left either by the fast-moving BCG as 
it plows through the inner ICM, or related to an observed sloshing cold front in the
inner part of the cluster 
\citep{Markevitch2001}. 
On even smaller scales, ALMA reveals
two massive molecular gas filaments that trace the edges of the inner radio bubbles within 
10 kpc of the AGN
\citep{Russell2017_A1795}.
These filaments exhibit a smooth velocity gradient spanning 450 km/s,
and they are thought to result from
low-entropy X-ray gas that has been 
entrained by the expanding bubbles,
become thermally unstable, and cooled.
A recent study of stellar and nebular emission line kinematics suggests a similar range 
of line-of-sight velocities for these
populations \citep{Tamhane2023_A1795}. 

In this paper we present XRISM/Resolve
observations of the central and northern
region of A1795. 
We focus on the ICM kinematics and
resonant scattering of the cool core and
regions immediately outside of the cool core.
Throughout the paper, we adopt a $\Lambda$CDM 
cosmology with
\( H_0 = 70 \ \, \text{km s}^{-1} \, \text{Mpc}^{-1} \),
\( \Omega_m = 0.3 \), 
and \( \Omega_\Lambda = 0.7 \).
At the redshift of A1795 
(\( z = 0.062 \)), 1$\arcmin$ 
corresponds to 69 kpc. 
{ All reported redshifts and velocities
have been corrected to the solar system
barycenter.}
We also adopt \citet{Lodders09}
proto-solar abundance table.
Unless otherwise noted, 
all uncertainties are 
reported at the
1\( \sigma \) (68\%) 
confidence level.

\section{Data Reduction}
XRISM observed Abell~1795 through
two pointings: one central pointing on 
2025 January 22--27 for 226 ks 
(OBSID 201087010) and one northern
pointing on 2025 January 20--22
for 113 ks (OBSID 201088010). 
Table \ref{tab:obs_log} shows 
a complete observation log.
The data were reprocessed 
with the \texttt{xapipeline} tool in HEASoft v6.36 \citep{2014ascl.soft08004N}, adopting the most recent calibration database 
(CalDB v20250915) for Resolve. Following the recommendations in the XRISM ABC Guide v1.0\footnote{\url{https://heasarc.gsfc.nasa.gov/docs/xrism/analysis/abc_guide/xrism_abc.html}}, we screened the event data for cross-talk events using the specified \texttt{RISE_TIME}, \texttt{DERIV_MAX}, and \texttt{STATUS} column filtering criteria.

\begin{table*}
\caption{XRISM Observation logs\label{tab:obs_log}}   
\begin{center}
\setlength{\tabcolsep}{6pt}
\begin{tabular}{cccccc}
Pointing & Observation ID & RA & Decl. & Observation Date & Exposure\\
& & (deg) & (deg) & & (ks)\\
\hline
\hline
Central & 201087010 & 207.21837 & 26.59235 & 2025-01-22--2025-01-27 & 225.8\\
North 1 & 201088010 & 207.21804 & 26.64328 & 2025-01-20--2025-01-22 & 113.3\\
\hline
\end{tabular}
\end{center}
\end{table*}

For each observation, spectra 
were extracted from the full array and
sub-arrays, excluding Pixel 27,
which has historically exhibited anomalous gain
shifts not aligned with the timing
of the filter wheel $^{55}$Fe gain fiducial exposures. 
Pixel 11 also occasionally displays similar abrupt scale variations,
therefore it is also excluded from our analysis
\citep{A2029_Sarkar}.
Our analysis includes only high-resolution primary (Hp; \texttt{ITYPE=0}) events,
which comprise more than 99\% of the 
2--10 keV event population in each observation.
For background, 
a non-X-ray background (NXB) 
spectrum was generated for each observation using the
{\tt rslnxbgen} tool.
Instrumental responses were 
generated for each observation 
using {\tt rslmkrmf} with the most
up-to-date calibration database 
(CalDB v20250915). 
The redistribution matrix file 
(RMF) was scaled by the fraction of high-resolution primary (Hp)
events in the 2--10 keV
band—excluding low-resolution
secondary (Ls) events—to properly account for the exclusion 
of medium- and low-resolution events and ensure accurate
flux normalization.

\section{Spectral fitting}\label{sec:spectral_fitting}

Spectral fitting was performed using {\tt XSpec 12.15.1} \citep{1996ASPC..101...17A}.
{ We extracted spectra from the pixels in the central and northern
pointings that lie within each annulus with radii of 
$0'$–$0.75'$, $0.75'$–$1.5'$, $1.5'$–$3'$, and
$3'$–$4.5'$, as shown in Figure~\ref{fig:chandra_image}.}
Given the modest point-spread function 
of the Resolve X-ray mirror assembly \citep[Half Power Diameter $\sim 1.3'$,][]{2025PASJ...77S...1T}, 
spatial spectral mixing (SSM) was taken into 
account following \citet{A2029_Sarkar}.
{ We used four annuli regions,
as shown in Figure~\ref{fig:chandra_image}, to model the ICM emission corresponding to
each extracted Resolve spectrum.} 
{ We generated ancillary response files (ARFs) with {\tt xaarfgen} in image mode, adopting the
2--10 keV Chandra image as the input 
sky brightness distribution. 
The {\tt xaarfgen} task internally calls {\tt xrtraytrace} to ray-trace photons through the X-ray Mirror Assembly and {\tt xaxmaarfgen} to compute the net effective area for the selected 
Resolve detector regions, including detector efficiencies. 
To account for the broad Resolve point-spread function and the resulting spatial mixing, 
we computed ARFs for each pairwise combination of input and extraction regions shown in Figure~\ref{fig:chandra_image}. 
These region-dependent ARFs were then used in XSPEC to fold each model component through the appropriate response. For more details on
{\tt xrtraytrace} tool, we refer readers to XRISM abc guide\footnote{\url{heasarc.gsfc.nasa.gov/docs/xrism/analysis/quickstart}}.}

\begin{figure*}
    \hspace{-0.9in}
 \includegraphics[width=1.2\textwidth]{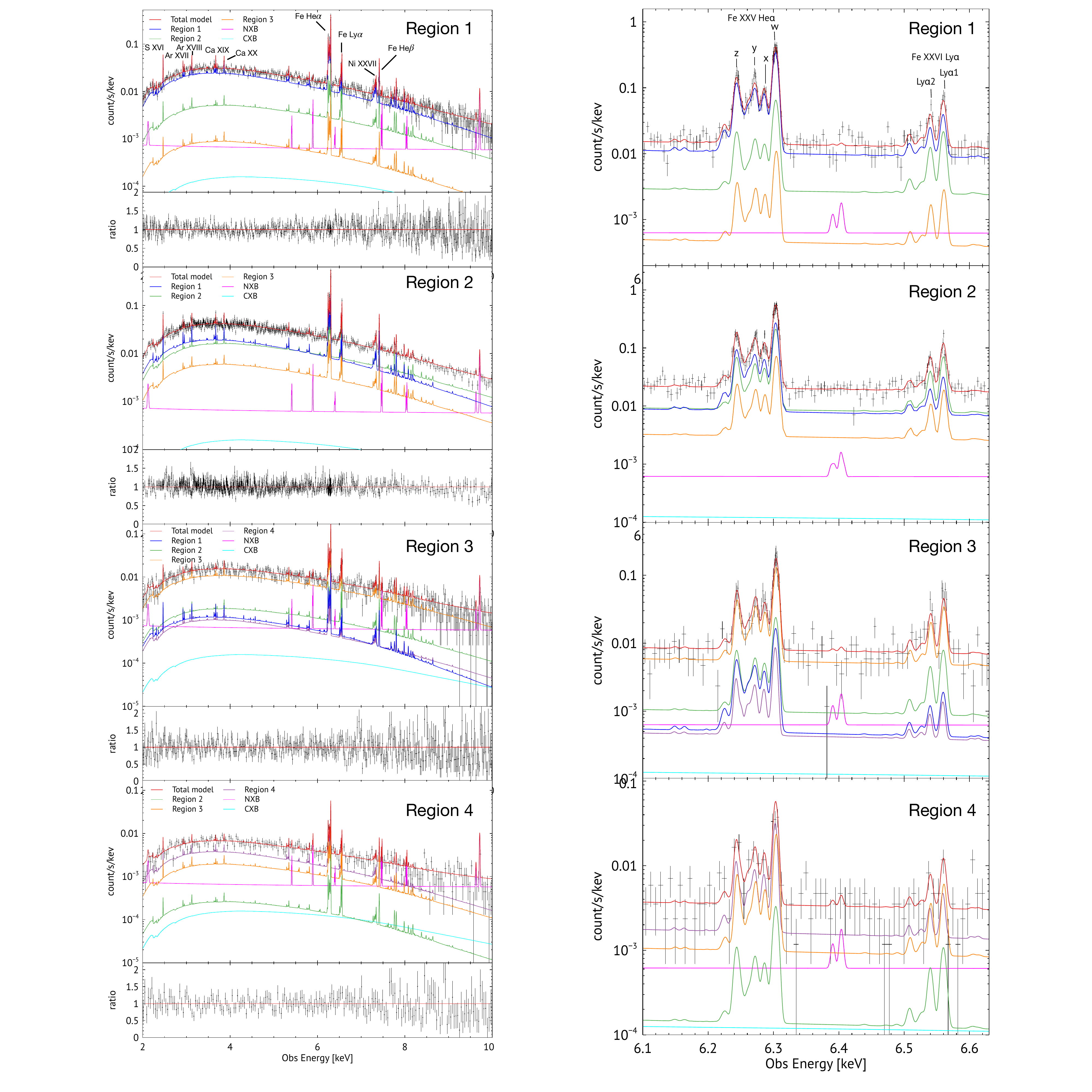}
    \caption{Left panels: XRISM/Resolve sub-array { counts} spectra are shown along with
    the best-fit 1T models after accounting
    for Spatial-Spectral Mixing. 
    In all four panels, the best-fit model
    components are { folded through the relevant instrumental response and plotted} as follows: red,
    blue, green, orange, purple, magenta, and cyan
    for the total model, region 1 (0--0.75$'$), region 2 (0.75$'$--1.5$'$),
    region 3 (1.5$'$--3$'$), region 4 (3$'$--4.5$'$), NXB, and CXB, respectively.
    Bottom panels in all four spectra represent
    ratios between data/model.
    Right panels: Same as left but zoomed into 
    $\fexxv$-He$\alpha$ and $\fexxvi$-Ly$\alpha$ line complexes.}
\label{fig:cen_resolve_spec}
\end{figure*}

\begin{figure*}
    \centering
 \includegraphics[width=1.05\textwidth]{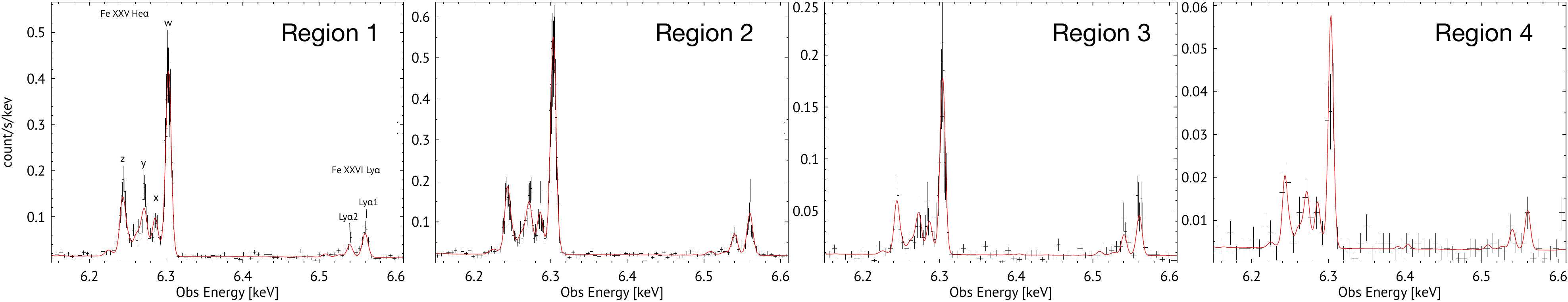}
    \caption{XRISM/Resolve sub-array {counts} spectra are shown together with the best-fit total 1T models, as in Figure~\ref{fig:cen_resolve_spec}, but with linear y-axes and zoomed in on the Fe-band region.}
\label{fig:cen_resolve_spec_zoomed_linear}
\end{figure*}

The spectrum from each region was fitted with an absorbed single-temperature (1T) {\tt TBABS $\times$ BAPEC} model.
{ We note that {\tt BAPEC} model accounts
for the thermal broadening.}
For each region, the temperature, 
abundance, redshift, velocity dispersion ($\sigma_v$), 
and normalization were allowed to vary freely.
The hydrogen column density, { $N_{H}$}, 
was fixed to the Galactic value of $1.2 \times 10^{20}$~cm$^{-2}$ \citep{2016A&A...594A.116H}.
In addition, the spectrum from each region was 
fitted with an absorbed two-temperature (2T) 
{\tt TBABS $\times$ (BAPEC+BAPEC)} model.
The non–X-ray background (NXB) was modeled by
fitting the NXB spectrum with a power law
and multiple Gaussian components representing instrumental 
lines, 
simultaneously with the source emission.
Similarly, the unresolved Cosmic X-ray Background 
(CXB) was modeled using a power law with a photon 
index of $\Gamma = 1.41$ and 
a scaled-normalization adopted from Suzaku measurements \citep{2009PASJ...61.1117B}.
We ignored the effect of Galactic foreground
components since our spectral fitting is limited
to $>2$ keV.
The best-fit parameters and their confidence 
intervals for each region were obtained by 
minimizing the C-statistic \citep{1979ApJ...228..939C}.
Table~\ref{tab:best_fit_param} lists the
best-fit parameters obtained from the
1T and 2T fits, together with their 1$\sigma$ 
uncertainties.
Figure~\ref{fig:cen_resolve_spec} shows the
best-fit 1T models along with the 
Resolve spectra for each region.
{ Figure~\ref{fig:cen_resolve_spec_zoomed_linear} shows the zoomed in version with the linear y-axes.}

\begin{table*}
\caption{Best-fit parameters from Resolve spectral fitting\label{tab:best_fit_param}}   
\begin{center}
\footnotesize
\setlength{\tabcolsep}{3pt}
\begin{tabular}{ccccccccc}
Model & Radius & $kT$ & Abun & Redshift$^{\ast}$ & $\sigma_v$ & $v_{\rm bulk}^{\ddagger}$ & norm$^{\dagger}$ & C-stat/d.o.f\\
& & (keV) & (solar) & & (km/s) & (km/s) & ($10^{12}$ cm$^{-5}$) & \\
\hline
\hline
1T & $0-0.75'$ & 3.92 $\pm$ 0.07 & 0.58 $\pm$ 0.03 & 0.06308 $\pm$ 0.00004 & 114 $\pm$ 11 & $22 \pm 12$ & 2.2 $\pm$ 0.1 & 17145/15999\\

 & $0.75'-1.5'$ & 6.28 $\pm$ 0.27 & 0.61 $\pm$ 0.06 & 0.06302 $\pm$ 0.00007 & 115 $\pm$ 23 & $7 \pm 21$ & 1.1 $\pm$ 0.1 & 17121/15999\\

 & $1.5'-3'$ & 5.53 $\pm$ 0.22 & 0.47 $\pm$ 0.05 & 0.06289 $\pm$ 0.00007 & 86 $\pm$ 25 & $-32 \pm 21$ & 1.9 $\pm$ 0.1 & 14122/15999\\

 & $3'-4.5'$ & 4.23 $\pm$ 0.37 & 0.28 $\pm$ 0.06 & 0.06317 $\pm$ 0.0002 & 68 $\pm$ 39 & $47 \pm 60$ & 1.2 $\pm$ 0.1 & 10594/15999\\

\hline
\hline
2T & $0-0.75'$ & 2.72 $\pm$ 0.67 & 0.63 $\pm$ 0.03 & 0.06308 $\pm$ 0.00004 & 187 $\pm$ 60 & 22 $\pm$ 12 & 0.7 $\pm$ 0.3 & 17138/15999\\
 & & 4.65 $\pm$ 0.35 & $-^{\dagger\dagger}$ & $-$ & 93 $\pm$ 24 & -- & 1.5 $\pm$ 0.5 & \\
 \hline
 & $0.75'-1.5'$ & 5.12 $\pm$ 0.71 & 0.54 $\pm$ 0.04 & 0.06303 $\pm$ 0.00009 & 87 $\pm$ 34 & 10 $\pm$ 27  & 0.9 $\pm$ 0.3 & 17115/15999\\
 & & 6.81 $\pm$ 1.22 & $-$ & $-$ & 175 $\pm$ 70 & --  & 0.4 $\pm$ 0.2 \\
 \hline
 & $1.5'-3'$ & 1.18 $\pm$ 0.76 & 0.55 $\pm$ 0.05 & 0.06289 $\pm$ 0.00007 & $-$ &  $-32\ \pm$ 21 & 0.4 $\pm$ 0.3 & 14117/15999\\
 & & 6.08 $\pm$ 0.34 & $-$ & $-$ & 90 $\pm$ 24 & -- & 1.6 $\pm$ 0.1 \\
 \hline
 & $3'-4.5'$ & 4.30 $\pm$ 0.38 & 0.29 $\pm$ 0.06 & 0.06317 $\pm$ 0.00017 & 63 $\pm$ 30 & 47 $\pm$ 51  & 1.2 $\pm$ 0.1 & 10588/15999\\
 & & $-$ & $-$ & $-$ & $-$ & & -- & $-$ \\
\hline
\hline
\end{tabular}
\end{center}
{ $^{\ast}$: Redshift is corrected for solar system barycenter.}
{ $^{\ddagger}$: Only statistical fitting uncertainties are shown; BCG redshift uncertainty and the instrumental systematic gain uncertainty are not included.}
$^\dagger$: {\tt BAPEC} normalizations correspond to
the annuli shown in Figure~\ref{fig:chandra_image}.
$^{\dagger\dagger}$: parameters are linked
between two {\tt BAPEC} temperature components for each
region.
\end{table*}

\section{Results and Discussion}

\subsection{ICM velocity structure}
\begin{figure} 
    \centering
 \includegraphics[width=0.45\textwidth]{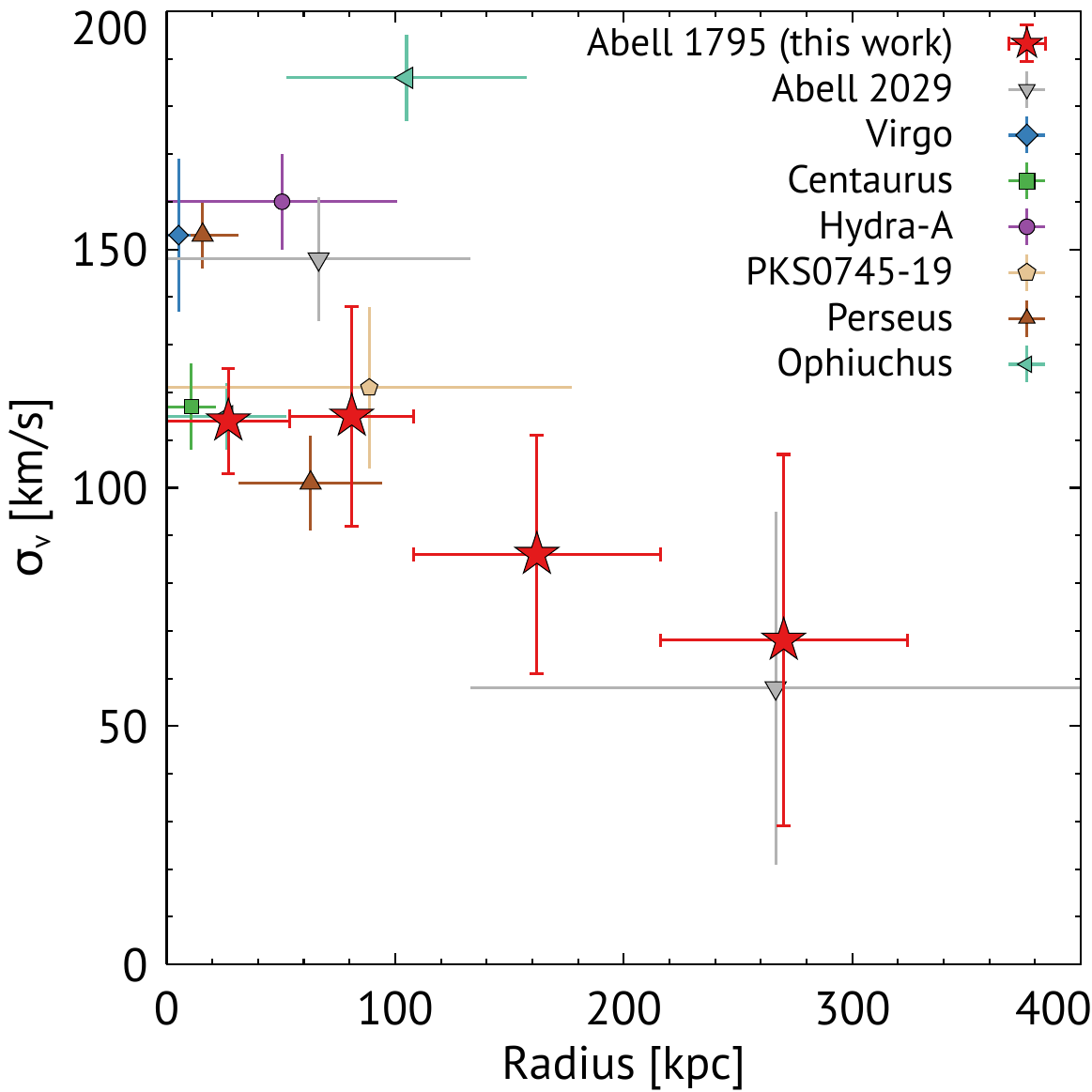}
    \caption{Radial profile of the line-of-sight velocity dispersion in A1795 (red star). Velocity dispersion measurements from other clusters are shown for comparison, including A2029 \citep{A2029_naomi,A2029_Sarkar}, Virgo \citep{XRISM2026_Virgo_paper1}, Centaurus \citep{2025Natur.638..365X}, Hydra A \citep{2025ApJ...990...42R}, PKS 0745–19 \citep{2026arXiv260316263T}, Perseus \citep{2026Natur.650..309T}, and Ophiuchus \citep{2025PASJ...77S.270F}.}
\label{fig:sigma}
\end{figure}

{ 
The velocity dispersion measurements from
Resolve spectra
are primarily constrained by the
Fe-K line complexes in the 6–8~keV energy band.
Figure~\ref{fig:sigma} presents the best-fit
velocity dispersion, $\sigma_{\rm v}$,
as a function of radius from the cluster center.
A1795 exhibits remarkably low velocity 
dispersion ($\sim$50–90~km~s$^{-1}$) just 
outside the central Resolve pointing 
($\gtrsim$100~kpc), comparable to values reported
for A2029 \citep{A2029_naomi,A2029_Sarkar}.
Within the central 100~kpc, 
the measured $\sigma_{\rm v}$ values 
($\sim$115~km~s$^{-1}$) are consistent with but on the low end of those 
observed in other clusters, including Ophiuchus \citep{2025PASJ...77S.270F}, Perseus
\citep{2016Natur.535..117H}, 
Centaurus \citep{2025Natur.638..365X},
and A2029 \citep{A2029_eric,A2029_naomi,A2029_Sarkar}. 
\citet{XRISM2025_cluster_velocities_vs_sims}
compared measured velocity dispersions of cool-core cluster centers with predictions from cosmological simulations, finding systematically lower velocities in the observed clusters. 
Our measurement of a low velocity dispersion in
A1795's core further strengthens the possible tension between
the simulations and measured gas kinematics.   
}

We adopt a redshift of $0.063001 \pm 0.000223$
for the BCG of A1795. 
This redshift corresponds to 
an observed 
barycentric systematic velocity of $cz = 18887\pm67$ km/s.
This value is derived from stellar population 
synthesis modeling of XSHOOTER spectra extracted from the central region of the BCG \citep{2023MNRAS.519.3338T}.
All bulk velocities reported in this paper
are measured relative to this systemic velocity using the expression
\begin{equation}
    v_{\rm bulk} = \frac{c(z-z_{\rm BCG})}{1+z_{\rm BCG}},
\end{equation}
where $z$ and $z_{\rm BCG}$ are the redshift 
of the hot ICM
measured by XRISM and the redshift of the BCG, respectively. 

\begin{figure}[h!]
    \centering
 \includegraphics[width=0.45\textwidth]{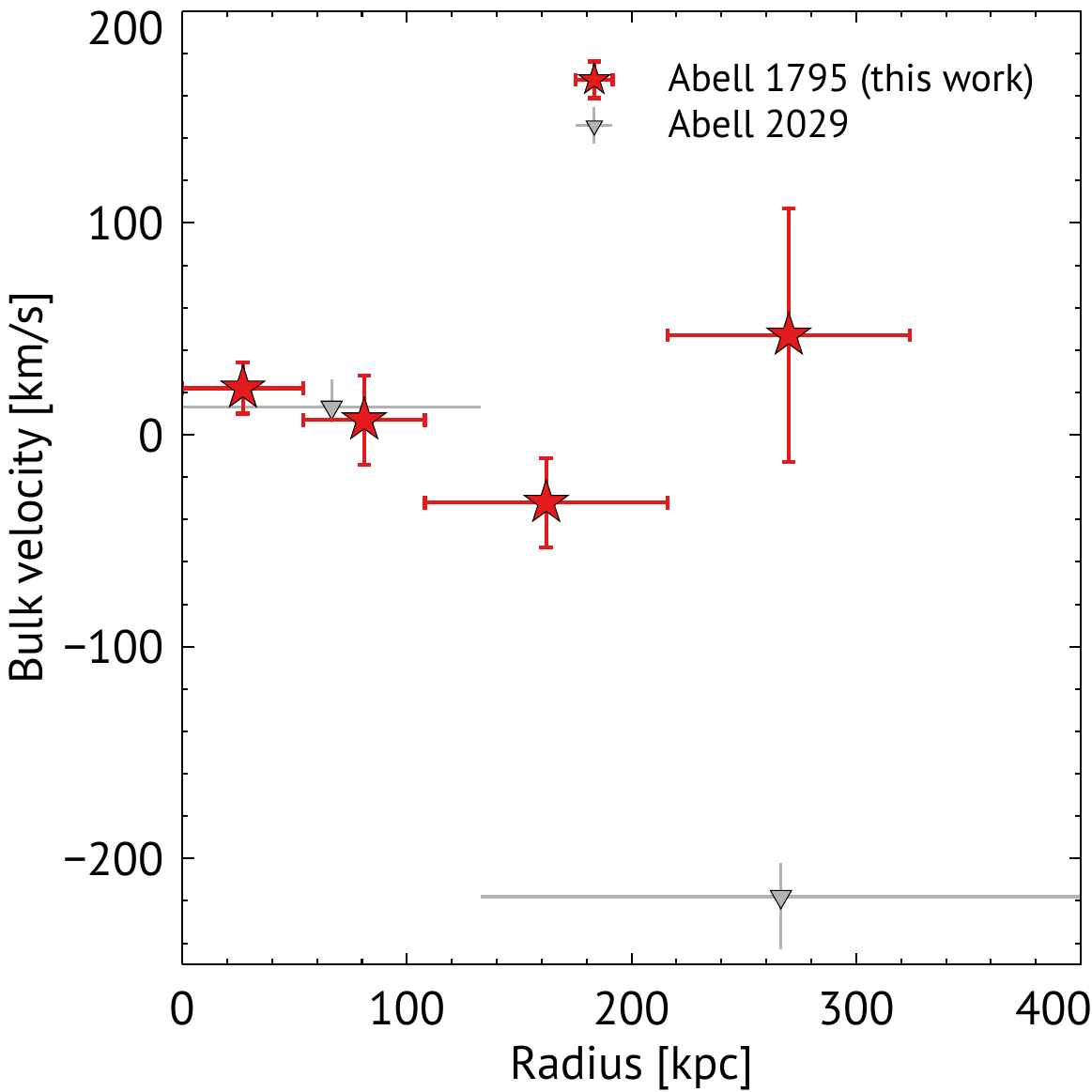}
    \caption{Bulk velocities in A1795 
    as a function of distance from the 
    cluster center (red star). 
    Grey inverted triangles indicate 
    the bulk velocities measured in the 
    A2029 cluster \citep{A2029_naomi,A2029_Sarkar}.}
\label{fig:bulk_vel}
\end{figure}

{ Figure \ref{fig:bulk_vel}
shows the radial profile of the ICM bulk velocity in
A1795 as a function of distance from the cluster center. 
Within the central $1.5'$ (113 kpc), the bulk velocity is
remarkably close to zero, suggesting no apparent
motion between the BCG and local ICM.
This result is in direct contrast to the
``cooling-wake'' scenario, in which the cool gas tail extending to the south is
produced as the central cluster galaxy oscillates through the cluster core 
\citep{2001MNRAS.321L..33F}.
If it is the case, a measurable velocity offset between the BCG and ICM would be expected.
Our results instead support the 
interpretation proposed by \citet{2026MNRAS.tmp..184V}, 
in which AGN-driven outflows uplift
low-entropy gas from the cluster center
to larger radii, where it eventually cools 
and condenses to form the observed tail.}

\subsection{Systematic uncertainties}

Systematic uncertainties can alter the reliability of the measured ICM bulk velocity and velocity dispersion. The most important systematic effects for XRISM Resolve observations arise from calibration uncertainties in the energy scale (or ``gain'') and the spectral line-spread function (LSF). The current in-flight gain calibration 1-$\sigma$ uncertainty is estimated to be $\pm0.3$ keV in the 5.4--9 keV band, and $\pm1$ keV below 5.4 keV, where high-fidelity calibration lines are not available on-orbit \citep{Eckart2025_JATIS}. The bulk velocities reported in the previous section are primarily constrained by the He-like and H-like Fe K complexes at 6.2--6.6 keV, therefore we adopt a systematic bulk velocity uncertainty of 0.3 keV $\approx$ 14 km\,s$^{-1}$. This is on the order of the lowest statistical error in the $v_{\rm bulk}$ measurements in Table \ref{tab:best_fit_param}. 

The LSF calibration uncertainty arises from two sources: uncertainty in the intrinsic LSF width for a given pixel, and the spread of gain solutions across all pixels included in the analysis, which can produce an effective broadening in the combined spectrum. The former is estimated from results and methods in \cite{Leutenegger2025} to be $\pm$0.16 eV, quoting 1-$\sigma$ uncertainty in the LSF width. When added in quadrature with the measured line widths, this represents a systematic uncertainty of $\lesssim4$ km\,s$^{-1}$ in all velocity dispersions reported in this work, significantly smaller than the statistical errors. For the latter effect of pixel-to-pixel gain smearing, we follow the lead of \cite{Simionescu2026_XRISM_Virgo_arms} and assume that this spread is the same as the overall gain uncertainty quoted above, $\pm0.3$ keV at 5.4--9 keV and $\pm1$ keV below 5.4 keV. This conservative assumption produces a systematic uncertainty of $\lesssim6$ km\,s$^{-1}$ for all reported velocity dispersions except those measured in the 2--4 keV band in Section \ref{sec:multiphase}, which have uncertainties of $\sim$55 km\,s$^{-1}$. These systematic uncertainties due to energy-related calibration do not alter the conclusions in this work.

Additional sources of systematic uncertainty include point-spread-function calibration and the use of SSM; the level of the background; the energy range; the amount of spectral binning; the presence of multi-temperature ICM; the presence of resonant scattering; and the versions of software and atomic databases employed. Several of these effects are explored in the following sections. For the others, specifically SSM, background, binning, and software/atomic database versions, we point to the results from \cite{A2029_naomi} and \cite{XRISM2026_Virgo_paper1}, who study these effects in detail for Resolve galaxy cluster observations and find no significant impact on their results. The observations and velocity structure results presented here for A1795 are very similar to those studies, therefore we conclude that these other systematic uncertainties should not alter the conclusions from this work.

\subsection{Non-thermal pressure support}

We follow the procedure of \cite{A2029_naomi} to infer the contribution of non-thermal pressure from the measured ICM velocity structure.
The ratio of non-thermal
pressure to the total pressure is estimated using
\begin{equation}
    \frac{P_{\rm NT}}{P_{\rm T}} = \frac{\mathcal{M}^2_{\rm 3D, eff}}{\mathcal{M}^2_{\rm 3D, eff} + \frac{3}{\gamma}},
    \label{eq:ntp}
\end{equation}
where $\mathcal{M}_{\rm 3D,eff}$ denotes the effective 
three-dimensional Mach number, 
computed by combining the contributions 
from both the velocity dispersion and the 
bulk velocity components, as
\begin{equation}
    \mathcal{M}_{\rm 3D, eff} = \frac{\sigma_{v,\rm eff}}{c_s} = 
    \frac{\sqrt{3\sigma_{v}^2 + v^2_{\rm bulk}}}{c_s}\, ,
    \label{eq:mach}
\end{equation}
where $c_s$ is the sound speed, estimated
as $c_s = \sqrt{\frac{\gamma k_B T}{\mu m_{p}}}$. Here, $\gamma=5/3$,
$k_B T$ is the measured X-ray temperature,
$\mu$ is the mean molecular weight, and
$m_p$ is the proton mass.
We make two important assumptions in Eqs. \ref{eq:ntp} and \ref{eq:mach}. First, we assume that all non-thermal pressure arises from turbulence, and we ignore other possible contributions like magnetic fields and cosmic rays \citep[e.g.,][]{ettori2022}. Second, we assume the measured velocity dispersion is due fully to isotropic turbulence, and we include the line-of-sight bulk velocity in our estimate of the three-dimensional Mach number. These assumptions are consistent with recent XRISM studies \citep[][e.g.,]{A2029_naomi}. We note that $\mathcal{M}_{\rm 3D,eff}$ ranges from 0.1 to 0.2 in the regions of our observation, indicating sub-sonic ICM motions.

\begin{figure}[h!]
    \centering
\includegraphics[width=0.45\textwidth]{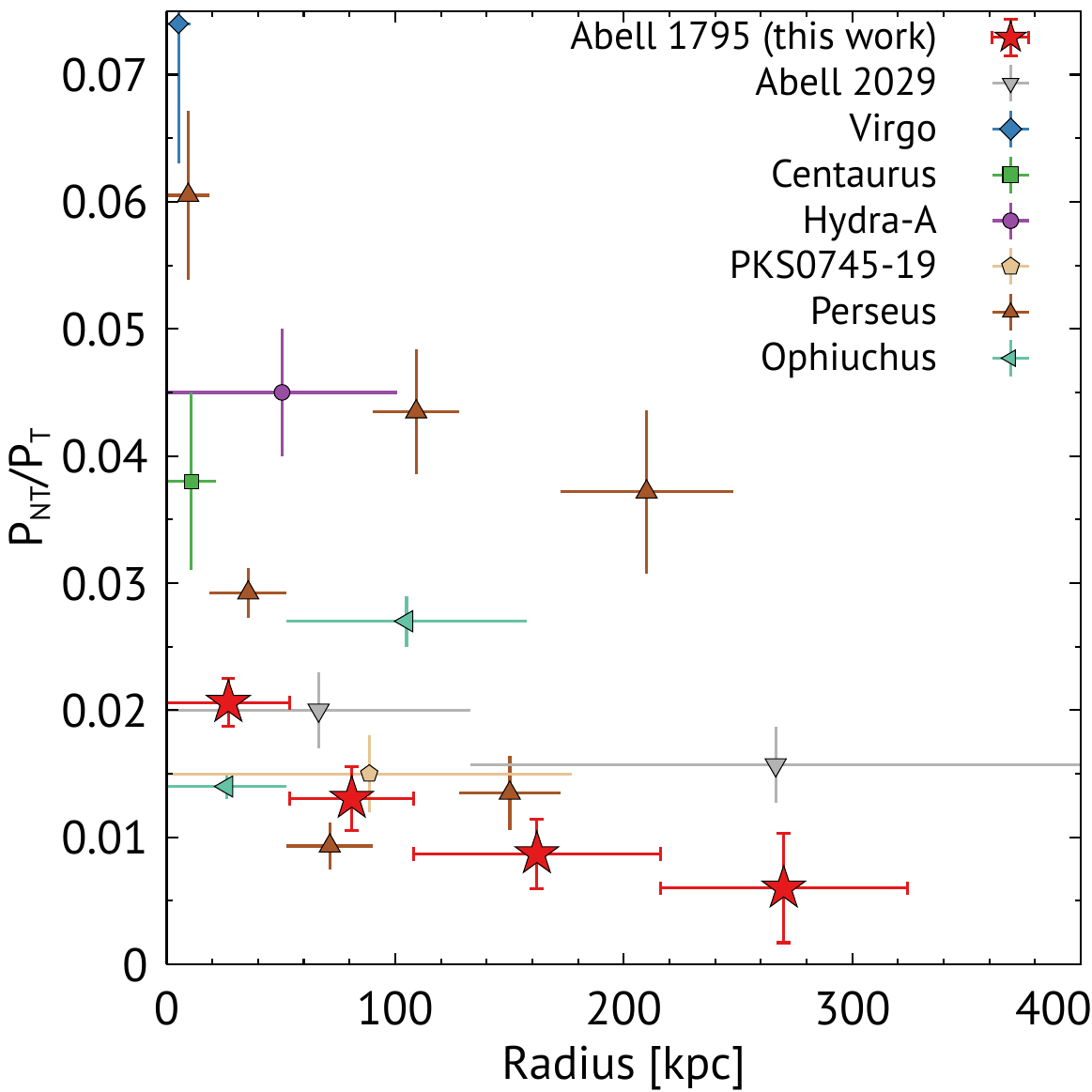}
    \caption{Radial profile of the
    non-thermal pressure fraction in A1795 measured from the cluster center 
    (red star). 
    Other data points show the
    non-thermal pressure fractions measured
    in other clusters, 
    as indicated in the legend.}
\label{fig:non_thermal_pres}
\end{figure}

Figure~\ref{fig:non_thermal_pres} shows the radial profile of the derived non-thermal–to–total pressure ratio, $P_{\rm NT}/P_{\rm T}$, in A1795.
Beyond $\sim$200~kpc from the cluster center, 
A1795 exhibits a remarkably low non-thermal pressure fraction ($\sim0.6\%$) compared to other clusters observed
with XRISM, indicating a quiescent
ICM just outside the cluster center.
This is the lowest non-thermal pressure fraction 
so far recorded in a cluster as measured by XRISM
far (at $\sim$300 kpc) from cluster core.
\citet{A2029_naomi} found a $P_{\rm NT}/P_{\rm T}$
$\sim$ $0.9\pm0.7$\% near R$_{2500}$ of 
A2029, claiming such low non-thermal pressure is rare for a relaxed cluster
and does not have a counterpart in the simulated
cluster sample \citep{XRISM2025_cluster_velocities_vs_sims}.
Our measurements of $P_{\rm NT}/P_{\rm T}$ in A1795 suggest that this is not rare. 
The non-thermal pressure may have a decreasing
trend for relaxed cool-core clusters reaching
below 1\% near R$_{2500}$.
Within 200~kpc, the non-thermal pressure fraction in 
A1795 ($\sim$1--2\%) is comparable to that measured
in A2029 \citep{A2029_eric,A2029_naomi}, Perseus
\citep{2016Natur.535..117H}, and PKS0745-19
\citep{2026arXiv260316263T}.

{ We compare our results with
that of previous indirect velocity measurements.
\citet{Heinrich2024} examined X-ray surface brightness fluctuations within R$_{2500}$ for a 
sample of luminous galaxy clusters observed with
Chandra. They derived power spectra
of density fluctuations and translated these
into velocity power spectra under the assumptions of
proportionality between two, adopting a calibration of the proportionality factor for clusters
in different dynamical states from \citet{2023MNRAS.520.5157Z}.
This analysis derived $P_{\rm NT}/P_{\rm T}$
values of $\sim$0.7--7\% within the central R$_{2500}$ of clusters similar to A1795.
We measure $P_{\rm NT}/P_{\rm T}$
between 2.1$\pm$0.2\% and
0.6$\pm$0.4\% within 0--0.5 R$_{2500}$ (0--320 kpc)
of A1795's center, with a median value of $\sim$1\%. These results are 
very close to the $P_{\rm NT}/P_{\rm T}$
derived from indirect measurements.} 

{ If the non-thermal pressure fraction in A1795
follows
the decreasing trend up to R$_{2500}$ (as shown in Figure~\ref{fig:non_thermal_pres}), it
does not significantly bias cluster mass measurements. However, our conclusion 
relies on $P_{\rm NT}/P_{\rm T}$ measurements along one azimuthal direction.
Simulations of cluster cores with prominent gas sloshing suggest that ICM gas 
velocities can be highly anisotropic,
and assuming isotropy can lead to underestimation
of the overall gas velocity \citep{2004ApJ...612L...9F}. 
The ICM of A1795 is remarkably smooth
just outside the central $\sim$100 kpc, showing
no prominent X-ray substructures, 
implying low non-thermal pressure at larger radii
in all directions. 
However, such confirmation needs future observations
sampling multiple directions of A1795.}

\subsection{Turbulent Heating}

A phenomenological turbulent heating model was fit to the X-ray luminosity profile of the cluster, where the heating power:
\begin{equation}
  P_{\rm turb} \simeq {3 \over 2} M(r){\sigma_v^3 \over l} 
  \label{eq:pturb}
\end{equation}
is assumed to balance cooling from the emission of X-rays.
$M(r)$ is the atmospheric gas mass subtended by the image within the radius that the velocity dispersion $\sigma_v$ is measured by XRISM. The gas mass $M(r)$ and X-ray luminosity $L_{\rm x}(r)$ are taken from the catalog of cluster thermodynamic profiles discussed by \cite{2018Pulido} and \cite{2017Hogan}. The turbulent dissipation timescale is assumed to be $\sim l/\sigma_v$. 
The energy injection scale $l$ cannot be measured by XRISM but is expected to be on the order of the X-ray bubbles and jets imparting energy into the X-ray atmosphere.  Those scales are typically a few to a few tens of kpc, which is below XRISM's spatial resolution. In general, $\sigma_v$ and $l$ must be measured on the same scales.  Instead, Eq. \ref{eq:pturb} is used to evaluate the radial range of $l$ required to instantaneously balance radiative losses, constrained by the observed velocity dispersion. 

If turbulent dissipation is the primary mechanism offsetting the cooling flow, it must operate over the cooling radius $\sim 130 ~\rm kpc$ \citep{2006Rafferty}.  The inner two velocity dispersion measurements roughly encompass this region.  The inner 60 kpc assumes $\sigma_v=114~\rm km~s^{-1}$ and from 60 kpc to the cooling radius $\sigma_v=115~\rm km~s^{-1}$ is assumed. Anchored by the measured velocity dispersions, $l$ was allowed to vary with radius so that at all radii $P_{\rm turb} \simeq L_{\rm x}(r)$. A radial trend in $l$ lying between $\simeq 2$ kpc in the inner 25 kpc rising to $\simeq 15$ kpc at the cooling radius would be required to fit the luminosity profile. This equilibrium model between energy injection, presumably by the jets and bubbles, and heating through a turbulent cascade in turn implies a timescale of $\tau \sim l/\sigma_v \sim 2\times 10^7 ~\rm yr$ in the inner 25 kpc and rising to $\sim 10^8~\rm yr$ at the cooling radius.
A complete description of this approach is given in \citet{2025ApJ...990...42R} and \citet{2026arXiv260419607M}.

{ If the turbulence energy propagates at
$v\sim114$ km/s, it would take $\sim1.1\times10^{9}$ yr
to reach the cooling radius, roughly 10 times
longer than the time scale needed for 
turbulent energy dissipation as heat
required to offset 
cooling.  However, turbulence diffuses much more slowly than the turbulent eddy speed (McNamara et al., in preparation),
which implies turbulent energy is dissipated 
locally, well before it can be transported
to larger radii. Turbulence is probably an inefficient mechanism for
redistributing AGN injected energy (via jets/bubbles) throughout the cooling volume.
Instead, these results support a scenario in which
heating must be spatially distributed, with
energy deposited in situ across the cooling 
region.
For turbulence to offset cooling 
entirely, it must propagate at a speed of
$\sim$1300 km s$^{-1}$, assuming cooling time
varies linearly with radius, which is roughly a factor of 10 higher than $\sigma_v$.
}

{ 
Our measurements indicate turbulent heating
alone struggles to offset cooling in A1795's
core. 
Similar conclusions have been drawn for Hydra-A,
Virgo, and Ophiuchus in previous
XRISM studies, suggesting additional heating channels, such as weak shocks and cosmic ray, are
required \citep[e.g.,][]{XRISM2026_Virgo_paper1,2025PASJ...77S.270F,2025ApJ...990...42R}. 
}

\subsection{Multi-phase ICM} 
\label{sec:multiphase}

We examine the presence of multiphase gas in A1795 by fitting the Resolve spectrum from each region with a 
two-temperature collisional ionization equilibrium (2T CIE) model, 
{\tt BVAPEC + BVAPEC}.
We allow the temperature ($kT$), 
velocity dispersion ($\sigma_v$),
and normalization of each component 
to vary independently,
while accounting for SSM using the same fitting procedure described in 
Section~\ref{sec:spectral_fitting}. 
The best-fit parameters from the 2T
fits for each region are listed in 
Table~\ref{tab:best_fit_param}.

Within the central 0.75$\arcmin$
($\sim$50 kpc), 
we detect two distinct temperature components:
a cooler component at $\sim$2.7 keV and a hotter component at $\sim$4.6 keV. 
This is consistent with the results of
\citet{2001A&A...365L..87T}, 
who found similar ICM temperatures within
the central $2'$ using XMM-Newton.
The cooler component shows a LOS
velocity dispersion that is higher by $\Delta\sigma_v \approx 90$ km/s
compared to the hotter component.
A similarly higher velocity dispersion 
is also observed in the 
0.75$\arcmin$–1.5$\arcmin$ ($\sim$50–103 kpc) region, 
with $\Delta\sigma_v = 88$ km/s,
although the two components have consistent temperatures. 
These results indicate the presence 
of multiphase gas in the central
regions of A1795.
In the 1.5$'$–3$'$ (103–206 kpc) region, 
we obtain a cooler temperature component 
of 1.2 keV; however,
the velocity dispersions of the 
individual components cannot
be constrained. 
For the 3$'$–4.5$'$ (206–310 kpc) region, the second temperature component is
not well constrained when left free and
is therefore tied to that of the first component.
In both outer annuli, the jointly fitted $\sigma_v$ values are consistent with
that of obtained from the 1T fit.

To confirm the presence of multi-phase gas
at the A1795 core, we re-fit 
the Resolve spectra from the $0-0.75'$ and 
$0.75'-1.5'$ annuli in two separate energy bins:
2--4 keV and 6--7 keV.
The motivation behind this is that the
2--4 keV energy band is more sensitive to 
cooler gas, while the 
6--7 keV band traces hotter 
gas, in both cases compared to
broadband 2--10 keV fitting.
The 
resulting best-fit parameters for
both regions are listed in 
Table \ref{tab:multi-range}. 
Within the central $0.75'$, 
the 2--4 keV band yields significantly lower
temperatures and metallicities
than those obtained from the 6--7 keV band.
The cooler gas component also has a higher
LOS velocity dispersion
than the hotter component.
In the $0.75'-1.5'$ region, both energy
bands yield
consistent temperatures and metallicities.
Although we are able to constrain 
$\sigma_v$ of $\sim100$ km/s from the 6--7 keV 
fit,
we cannot constrain 
$\sigma_v$ from the 2--4 keV fit due to low
statistical quality, however a 
1$\sigma$ upper limit of 200 km/s is obtained. 

The split-band results are consistent with
the best-fit parameters obtained from 
the 2T broadband fitting 
for both regions. 
These results suggest the presence of
multiphase gas in the central region of A1795,
in line with previous Chandra and GMRT
observations \citep{2018A&A...618A.152K}. 
The central 80 kpc of A1795 is morphologically
disturbed, showing multiple X-ray cavities,
surface brightness edges, and a cool
gas tail extending to the south, as shown in
Figure \ref{fig:chandra_image}. 
These features result from 
multiple episodes of mechanical feedback
from the central AGN, which shock-heat
the ICM, inject turbulence \citep{2012ApJ...746...94G,2012NJPh...14e5023M},
and uplift cool gas from the center \citep{2026MNRAS.tmp..184V}, giving
rise to a multiphase ICM in the core of A1795.

\begin{table*}
\caption{Best-fit parameters for central two regions fitted in two spectra ranges: 2--4 keV and 6-7 keV\label{tab:multi-range}}   
\begin{center}
\footnotesize
\setlength{\tabcolsep}{4pt}
\begin{tabular}{ccccccccc}
Radius & $kT$ & Abun & Redshift & $\sigma_v$ & $v_{\rm bulk}$ & norm\\
& (keV) & (solar) & & (km/s) & (km/s) & ($10^{12}$ cm$^{-5}$) \\
\hline
\hline

 & & & Range: 2--4 keV & &  &  &  &\\
 \hline
$0-0.75'$ & 3.40 $\pm$ 0.20 & 0.47 $\pm$ 0.11 & 0.06279 $\pm$ 0.00022 & 189 $\pm$ 61 & $-60 \pm 66$ & 2.1 $\pm$ 0.1\\

$0.75'-1.5'$ & 4.98 $\pm$ 0.47 & 0.38 $\pm$ 0.17 & 0.063335 $\pm$ 0.00045 & $<200$ & $98 \pm 135$ & 1.8 $\pm$ 0.1\\
\hline
 & & & Range: 6--7 keV & &  &  &  &\\
 \hline
$0-0.75'$ & 4.62 $\pm$ 0.16 & 0.68 $\pm$ 0.05 & 0.06311 $\pm$ 0.00004 & 115 $\pm$ 12 & $30 \pm 12$ & 1.8 $\pm$ 0.1\\

$0.75'-1.5'$ & 5.57 $\pm$ 0.23 & 0.47 $\pm$ 0.05 & 0.06295 $\pm$ 0.00006 & 107 $\pm$ 19 & $-15 \pm 18$ & 1.7 $\pm$ 0.1\\
\hline
\hline
\end{tabular}
\end{center}
\end{table*}

\subsection{Resonant Scattering}
\begin{figure}
    \centering
    \includegraphics[width=0.48\textwidth]{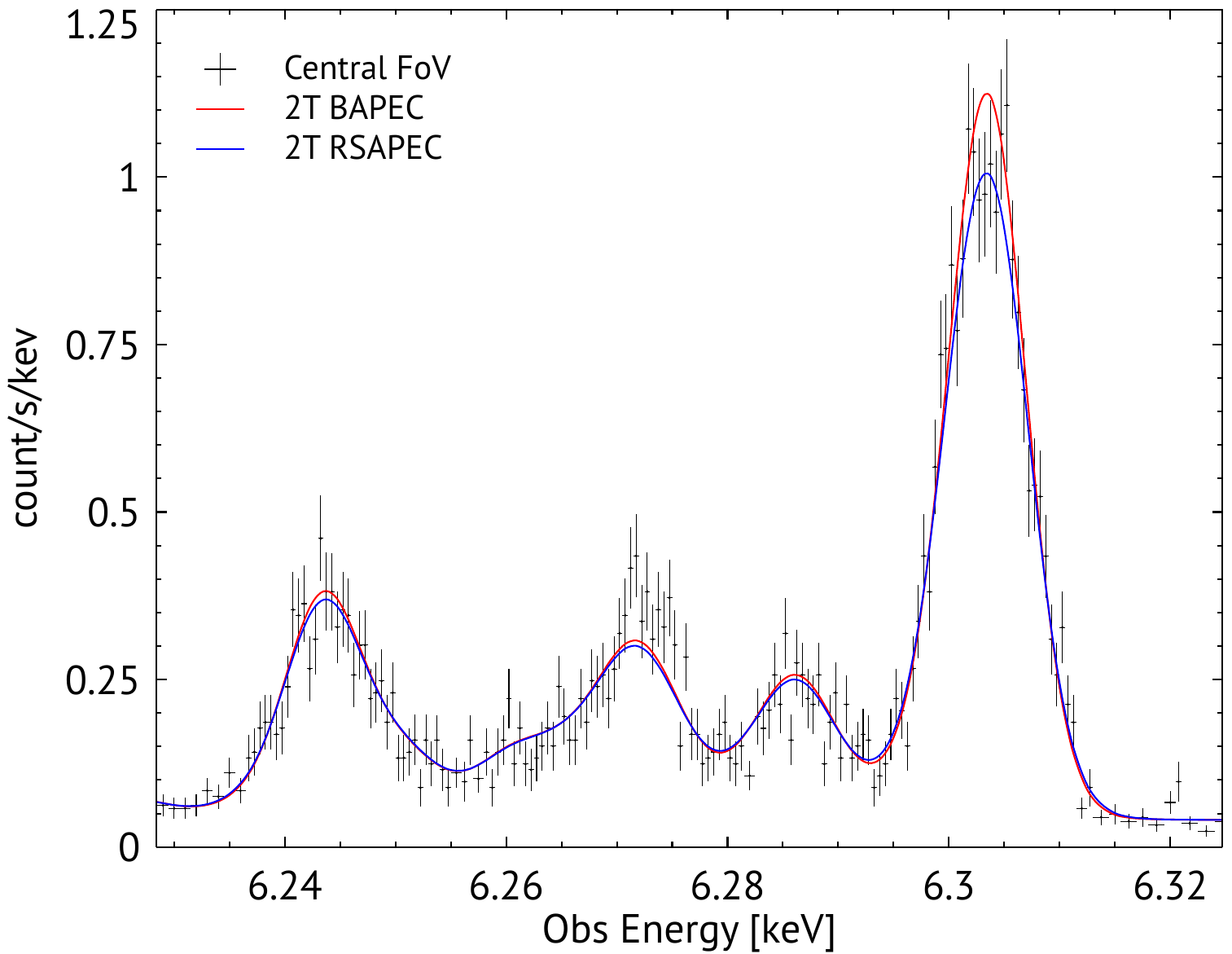}
    \caption{Resolve spectrum from the entire
    central pointing (black data points) fitted with single temperature
    velocity broadened CIE model without accounting 
    for resonant scattering (BAPEC; red)
    and with accounting resonant scattering
    (RSAPEC; blue). 
    Line flux of optically thick
    $\fexxv$ $w$ line is suppressed by a factor of
    $\sim$ 14\%.}
    \label{fig:resonant_scattering_spec}
\end{figure}

Line intensities of optically thick transitions
in the observed X-ray spectra can be reduced by resonant scattering,
in which line photons are absorbed and re-emitted in different directions, 
effectively scattering photons out of the line of sight and suppressing 
the observed line strength 
\citep[e.g.,][]{2004MNRAS.347...29C,2006MNRAS.370...63S,2023MNRAS.522.3665N}.
Resonant scattering effects are expected 
to be strongest in the central regions of clusters, 
where the gas density is highest.
Because the Resolve gate valve was closed during the observations,
the only potentially optically thick line 
resolved
in A1795 with XRISM at high significance 
is the $\fexxv\ w$ line at 6.7~keV.
We estimate the degree of resonant line
suppression in the central region of A1795
by fitting Resolve full-array spectra
extracted from the cluster core with a 
two-temperature {\tt RSAPEC} model 
\citep{2023ApJ...959..126C} in {\tt XSPEC}.
The {\tt RSAPEC} 
model provides a simple, single-step, 
and computationally efficient treatment of resonant scattering, 
and serves as an alternative to the {\tt BAPEC} model for modeling spectra from collisionally 
ionized plasma.
Figure~\ref{fig:resonant_scattering_spec} compares the best-fit collisional ionization equilibrium
(CIE) models before and after accounting for
resonant scattering, together with the observed Resolve spectra from the central 
region of A1795.

For both temperature components, 
we allow the temperature ($kT$), abundance, 
redshift, and normalization to vary freely, 
while the line-of-sight column density
(in the {\tt RSAPEC} model) is tied between
the two components.
Because resonant scattering suppression depends sensitively on velocity broadening, 
the velocity dispersion parameters, $\sigma_v$,
are also allowed to vary independently
for both components during the fitting.
Table~\ref{tab:rsapec_comp} summarizes the best-fit parameters obtained for
the two-temperature models.
For the {\tt RSAPEC} fit, 
we find best-fit temperatures of 
$3.86 \pm 0.19$~keV and $6.61 \pm 0.59$~keV, 
with corresponding velocity dispersions of 
$98 \pm 16$~km~s$^{-1}$ and 
$130 \pm 17$~km~s$^{-1}$. 
These values are consistent with those
derived from the two-temperature {\tt BAPEC} model, 
which does not account for resonant 
scattering effects.
By comparing the $\fexxv\ w$ line flux predicted
by the best-fit {\tt BAPEC} model with that from
the {\tt RSAPEC} fit,
we infer a resonant scattering suppression factor
of $\sim$14\%. 
As expected, the suppression is strongest
in the cluster core.
Resonant scattering effects can be even more 
pronounced in lower-energy lines in cooler
clusters such as A1795; 
however, Resolve lacks sensitivity below 2~keV 
due to the loss of low-energy response. 
Future X-ray microcalorimeter missions, 
such as NewAthena, 
will therefore be essential for detailed 
studies of resonant scattering in galaxy clusters.
We are unable to constrain resonant scattering 
in the northern region because of the low ICM surface brightness, 
and deeper observations will be required to improve these constraints.

\begin{table*}
\caption{Best-fit parameters for full-array 
central FoV fitted with 2T {\tt BAPEC} and
{\tt RSAPEC} models within 2--10 keV. \label{tab:rsapec_comp}}   
\begin{center}
\footnotesize
\setlength{\tabcolsep}{4pt}
\begin{tabular}{ccccccccc}
Model & $kT$ & Abun & Redshift & $\sigma_v$ & Column density & norm\\
& (keV) & (solar) & & (km/s) & (10$^{22}$ cm$^{-2}$) & ($10^{12}$ cm$^{-5}$) \\
\hline
{\tt BAPEC} & 4.62 $\pm$ 0.16 & 0.68 $\pm$ 0.05 & 0.06311 $\pm$ 0.00004 & 115 $\pm$ 12 & $-$ & 1.8 $\pm$ 0.1\\

& 5.57 $\pm$ 0.23 & 0.47 $\pm$ 0.05 & 0.06295 $\pm$ 0.00006 & 107 $\pm$ 19 & $-$ & 1.7 $\pm$ 0.1\\
\hline
{\tt RSAPEC} & 3.86 $\pm$ 0.19 & 0.59 $\pm$ 0.04 & 0.06303 $\pm$ 0.00006 & 98 $\pm$ 16 & 0.2 $\pm$ 0.1 & 2.4 $\pm$ 0.4\\

& 6.61 $\pm$ 0.59 & 0.59 $\pm$ 0.05 & 0.062304 $\pm$ 0.00006 & 130 $\pm$ 17 & $-^{\dagger}$ & 1.6 $\pm$ 0.4\\
\hline
\hline
\end{tabular}
\end{center}
$^\dagger$parameters are linked between the two temperature components
\end{table*}

\subsection{$\fexxv$ $y$ line anomaly}\label{sec:y_line_anomaly}
\begin{figure}
    \centering
    \includegraphics[width=0.48\textwidth]{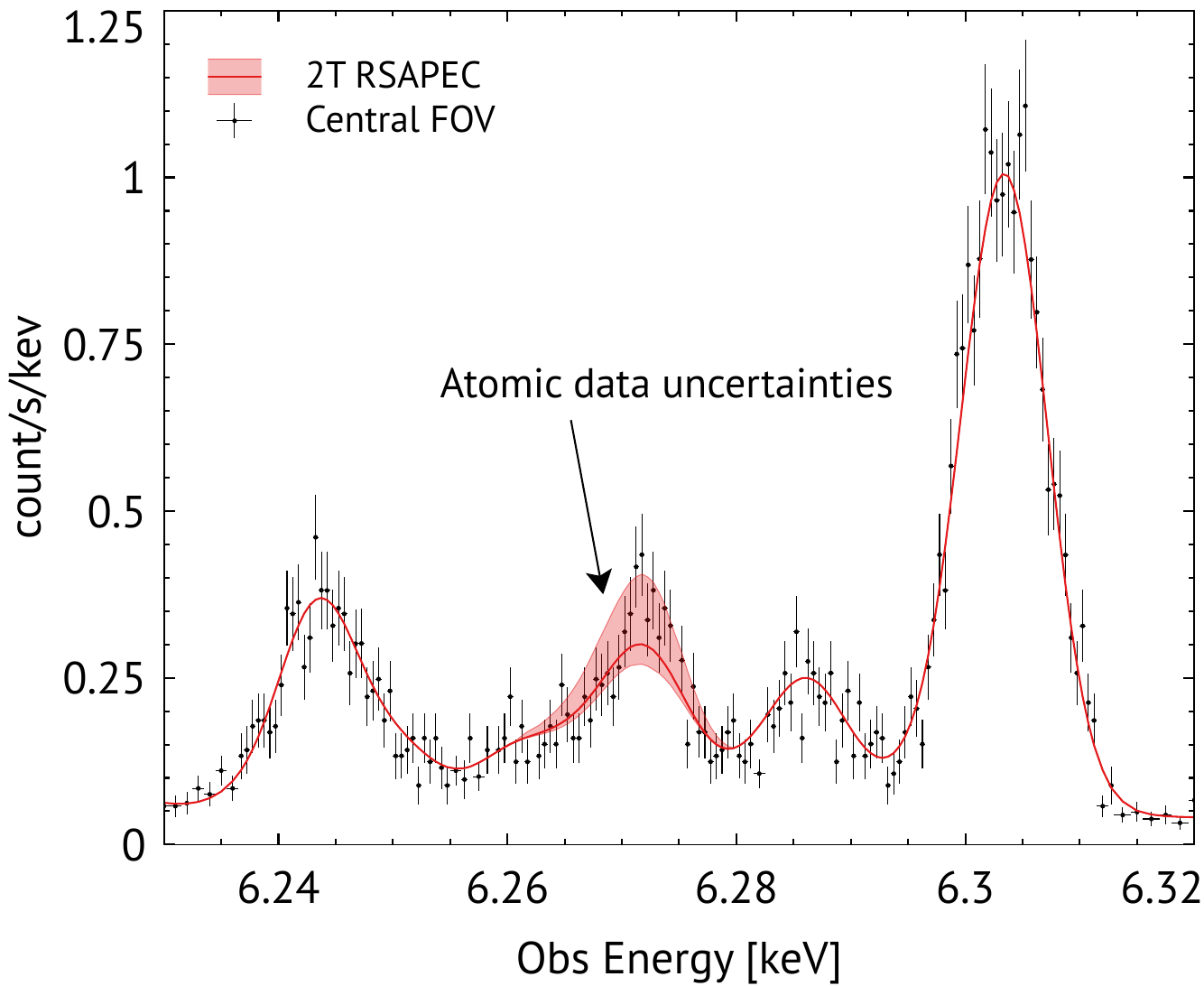}
    \caption{Resolve spectrum extracted from
    the central FOV (black data), 
    as shown in Figure 
    \ref{fig:chandra_image}, fitted with a 
    two-temperature CIE model 
    (red curve) accounting for
    resonant scattering of optically thick
    lines. Shaded region shows the variation
    of $\fexxv$-He$\alpha$ $y$-line emissivity
    due to atomic data uncertainties
    in the transition
    probability and collision strength.}
    \label{fig:y_line_anomaly}
\end{figure}

We identify a significant positive residual
in the $\fexxv$ He$\alpha$ $y$ 
intercombination line at 6.668~keV 
in the Resolve spectra of A1795 when 
fitted with a two-temperature collisional ionization equilibrium model that 
includes resonant scattering effects 
(2T {\tt RSAPEC}).
Figure~\ref{fig:y_line_anomaly} 
illustrates this anomaly in the full-array Resolve spectrum extracted
from the central region of A1795.
Such enhanced observed flux in the
$\fexxv$ He$\alpha$ $y$ line, 
compared to that predicted by the best-fit model, 
has also been reported in other clusters, 
including Ophiuchus
\citep{2025PASJ...77S.270F} and
Abell~2029 \citep{A2029_eric}.
We investigate possible origins of 
this line anomaly in A1795. 
In this work, we limit our analysis
to the central pointing, 
where the spectral quality is significantly higher than in the 
northern pointing.
We find that even after accounting for resonant
scattering and fitting with a 2T CIE model,
a substantial residual in the $y$ line persists, 
as shown in Figure~\ref{fig:y_line_anomaly}.

\citet{2024ApJ...962..192C} demonstrated
that the physical 
interpretation of X-ray spectra 
depends critically on the accuracy of
the atomic data used in spectral 
synthesis codes (see also \citealt{2018PASJ...70...12H}).
For collisionally ionized ICM plasma,
spectral line emissivities are 
particularly sensitive to uncertainties
in the Einstein $A$ coefficients and the collisional rate coefficients,
which represent the Maxwellian-averaged
collision cross sections ($q$ values).
We investigate whether the excess $y$-line
flux residual observed in A1795 can be 
explained by uncertainties in the
underlying atomic data. 
To do this, we use the {\tt variableapec}
module within pyatomdb \citep{atoms8030049},
as described in \citet{2024ApJ...962..192C}.
The {\tt variableapec} module allows
individual atomic parameters, 
such as Einstein \(A\) coefficients 
and collisional rate coefficients,
to be varied independently and 
computes the resulting changes in 
individual line emissivities.
The modified emissivities are then 
incorporated into {\tt XSPEC} by
updating the {\tt apecv3.1.2_line.fits} file.

To isolate the impact of atomic data
uncertainties on the $y$-line emissivity,
we vary only the Einstein $A$ coefficients
and collisional rate coefficients associated
with the $y$ transition.
Based on Table~3 of \citet{2018PASJ...70...12H}
and Table~1 of \citet{2024ApJ...962..192C},
we adopt uncertainties of 14\% in the Einstein \(A\) 
coefficients and 18\% in the collisional 
rate coefficients for the $y$-line. 
We then use {\tt variableapec} to quantify
the resulting changes in the $y$-line 
intensity using a Markov Chain Monte Carlo
chain with 
1000 realizations. Figure~\ref{fig:y_line_anomaly} 
shows the variation in the $y$-line emissivity
as the atomic parameters are varied within
these uncertainty ranges,
with the modified emissivities 
propagated through {\tt XSPEC}.
We find that some sets of Einstein 
$A$ coefficients and collisional rate coefficients increases
the predicted $y$-line emissivity 
(shade region in Figure~\ref{fig:y_line_anomaly}),
thereby reducing the discrepancy 
between the model and the observed spectrum.
However, we caution that this analysis 
provides only suggestive evidence that uncertainties
in the atomic data may contribute to 
the observed excess $y$-line flux, in
combination with the physical properties of the ICM. 
Additional high-quality Resolve cluster
spectra will be required to 
robustly determine the origin of the 
enhanced $y$-line emission.

\section{Summary}\label{sec:summary}
We present a total of $\sim$338 ks XRISM Resolve observations
of the A1795 galaxy cluster. Our main results are
summarized below.

\begin{itemize}
    \item We split two consecutive Resolve 
    pointings (central and North) into four
    regions and extracted spectra from 
    each region.
    Single-temperature fits reveal
    a gradient in line-of-sight velocity
    dispersion, decreasing from 114 km\,s$^{-1}$
    to 68 km\,s$^{-1}$ from the central
    to the outermost region at $\sim330$ kpc. 

    \item The measured 
    bulk velocities in the two central regions
    are low: $-4$ and $-10$ km\,s$^{-1}$, suggesting
    no significant relative motion between
    the BCG and ICM. 
    Because the innermost region includes the cool
    gas tail extending to south, such low bulk
    velocity directly contradict the ``cooling-wake''
    scenario for the origin of the tail. Instead,
    it supports the AGN-uplift origin.

    \item The non-thermal pressure fraction
    shows a decreasing trend with radius,
    with  
    $P_{\rm NT}/P_{\rm T}=0.019\pm0.002$ 
    in the core, declining to 
    a remarkably low value of $0.006\pm0.004$ at
    330 kpc from the cluster center. 
    Such low non-thermal pressure indicates that the
    ICM 
    in the northern region of A1795 is largely quiescent.

    \item { Our measurements suggest turbulent
    heating alone struggles to balance the cooling
    within the cooling radius ($\sim$130) of A1795.
    It supports a scenario in which
    heating must be spatially distributed, with
    energy deposited in situ across the cooling region.}

    \item We have identified two gas phases with 
    $\Delta\sigma_{v}\approx94$ km\,s$^{-1}$ and 88 km\,s$^{-1}$ 
    within the central $<0.75'$ and $0.75'-1.5'$ regions,
    respectively, based on a two-temperature fit and split
    energy band fit (2--4 keV and 6--7 keV).
    These results strongly suggest the presence
    of multiphase gas in the center of A1795.

    \item A $\sim$14\% 
    suppression of the optically-thick $\fexxv-w$
    line-flux in the central
    region of A1795 has been detected, suggesting resonant scattering of
    line-of-sight X-ray photons from the 
    cluster core to 
    larger radii.

    \item We have identified a significant positive
    residual in the observed flux of
    the $\fexxv-y$ line at 6.668 keV when
    fitted with CIE models. 
    Similar enhancements have been reported
    in other clusters,
    such as
    Ophiuchus and A2029. 
    We find that including atomic data
    uncertainties in the
    Einstein $A$ coefficient and 
    collisional rate of $y$-line reduces
    the discrepancy
    between the observation and the models. 
    Our analysis provides a suggestive evidence
    that underlying uncertainties in atomic data
    may (partially) contribute to the observed excess
    $y$-line flux.
    
\end{itemize}

\section*{Acknowledgement}
We thank the anonymous referee for their helpful comments.
We gratefully acknowledge the dedicated efforts of the many engineers and scientists whose hard work over the years made the XRISM mission possible.
AS acknowledges support from NASA grant 80NSSC26K0916 and 
EDM acknowledges support from NASA grants
80NSSC25K7538 and 80NSSC24K0678.

\bibliography{sample631}{} 
\bibliographystyle{aasjournalv7}

\end{document}